\documentclass{rsproca}

\usepackage{axodraw}
\begin{document}

\title{Multiple solutions in supersymmetry and the Higgs}

\author{
B. C. Allanach$^1$}

\address{$^{1}$DAMTP, CMS, Wilberforce Road, University of
  Cambridge, 
  Cambridge, CB3 0WA, United Kingdom}

\subject{Physics, Particle Physics, Mathematics}

\keywords{MSSM, technical hierarchy problem, constrained minimal
  supersymmetric standard model}

\corres{B. C. Allanach\\
\email{B.C.Allanach@damtp.cam.ac.uk}}

\begin{abstract}
Weak-scale supersymmetry is a well motivated, if speculative, 
theory beyond the 
Standard Model of particle physics. It solves the thorny issue of the Higgs
mass, namely: how can it be stable to quantum corrections, when they are
expected to be $10^{15}$ times bigger than its mass? The experimental signal
of the 
theory is the production and measurement of supersymmetric particles in the
Large Hadron Collider experiments. No such particles have been seen to date,
but hopes are high for the impending run in 2015. 
Searches for supersymmetric particles can be difficult to interpret. Here, we
shall discuss the fact that, even given a well defined model of supersymmetry
breaking with few parameters, there can be multiple solutions. These multiple
solutions are physically different, and could potentially mean that points in
parameter space have been ruled out by interpretations of LHC data when they
shouldn't have been. We shall review the multiple solutions and illustrate
their existence in a universal model of supersymmetry breaking.
\end{abstract}


\begin{fmtext}
\section{Introduction to Supersymmetry}
The recent discovery of the Higgs boson of mass 125-126 GeV at the Large
Hadron 
Collider 
experiments~\cite{ATLAS:2012oga,Chatrchyan:2012ufa} introduces a new problem: the technical hierarchy
problem. This problem concerns quantum corrections to the Higgs mass. As other
particles (which couple to the Higgs boson) fluctuate in and out of the
vacuum, they give a contribution to the Higgs mass squared of order
\begin{equation}
\delta m_h^2 = \frac{1}{16 \pi^2} M^2,
\end{equation}
where $M$ is the mass of the particles that are in the vacuum
fluctuations. 
\end{fmtext}

\maketitle

Thus, the dominant Higgs mass correction is expected to be from
the heaviest such mass scales. The technical hierarchy problem originates from
the observation that there are expected to be values that are many orders of
magnitude heavier than the Higgs mass. For example, we know of the existence
of the Planck scale, whose associated mass scale is $M_{Pl} \sim 10^{19}$
GeV. If this 
mass scale is (as we might expect) derived from microscopic propagating
degrees of freedom, they are then expected to
contribute to the Higgs mass with quantum corrections that are some 10$^{15}$
times higher than the measured mass. One could hope that several such
corrections 
cancel each other to 1 part in $10^{15}$, but many of us feel that this is 
unrealistic unless there is some underlying reason for the cancellation. 
\begin{figure}[!h]
\begin{center}
\begin{picture}(250,50)(0,30)
\DashLine(0,50)(33,50){5}
\ArrowArc(50,50)(16.5,0,180)
\ArrowArc(50,50)(16.5,180,360)
\DashLine(66,50)(100,50){5}
\Text(50,60)[c]{$F$}
\Text(50,40)[c]{$F$}
\Text(16.5,55)[c]{$h$}
\Text(83.5,55)[c]{$h$}
\Text(60,50)[c]{$i\lambda$}
\Text(40,50)[c]{$i\lambda$}
\Text(125,50)[c]{$+$}
\DashLine(150,25)(250,25){5}
\DashCArc(200,50)(25,0,360){5}
\Text(165,31)[c]{$h$}
\Text(235,31)[c]{$h$}
\Text(200,33)[c]{$i\xi$}
\Text(180,50)[c]{$\varphi$}
\Text(222,50)[c]{$\varphi$}
\end{picture}
\end{center}
\caption{Examples of Feynman diagrams that gives a large quantum correction to
  the   Higgs mass. The masses of the particles $F$ or $\varphi$ may be much
  larger than the   measured 
  Higgs boson mass. $\lambda$ and $\xi$ denote coupling strengths between the
  various fields. The first Feynman diagram shows a Higgs particle splitting
  up into a particle and an anti-particle, which recombine into a Higgs
  particle. The second diagram shows a Higgs particle interacting with
  $\varphi$ particles that are fluctuating out of the vacuum.} 
\label{hierProb}
\end{figure}
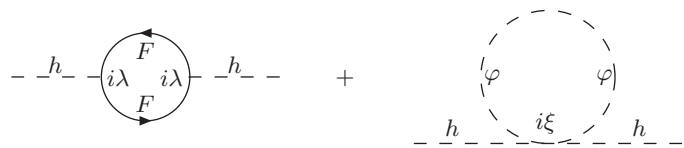

It must be said that these Planck scale masses are associated with gravity,
and a 
full quantum gravitational theory remains unverified by experiment. So one
possibility is that using the standard quantum field theory arguments about
vacuum fluctuations is simply wrong for some unknown reason
(for example, perhaps $M_{Pl}$ is a coupling constant that is not generated by
some microscopic degrees of freedom). However, there
are also other well-motivated extensions to the Standard Model of particle
physics where the forces are unified (grand unified theories). These still
have a huge mass scale associated with them of order  $10^{16}$ GeV and would
generate a correction to the Higgs mass that is much larger ($10^{12}$ times)
larger than its measured value.
In the Standard Model, it is only the Higgs boson that has this problem of
being sensitive to large quantum corrections. All of the particles other than
the Higgs boson are
protected by various symmetries: matter fields by the chiral symmetry of the
model, and the force-carrying gauge bosons are protected by the gauge symmetry
upon which 
the model is built. 
We emphasise however that there is technically no fine-tuning in the pure
Standard Model Higgs mass computation: there is no higher mass scale within
the Standard Model, because it does not include gravity or higher scales such
as those derived from grand unified theories. However, we shall proceed on the
basis that the hierarchy problem is pointing us in an interesting direction if
we take it seriously as a problem, expecting that gravitational degrees of
freedom will induce a huge quantum correction to the Higgs mass. 

If we examine the correction to the Higgs mass shown in
Fig.~\ref{hierProb}, we notice an interesting fact: the large corrections have
an different sign for the first contribution compared to the second. In fact,
this is 
a property of quantum field theory: fermions give a {\em negative}~ sign,
whereas bosons give a positive sign. 
Supersymmetry provides a
mathematical reason for a large cancellation between the two diagrams, by
imposing a symmetry on the quantum field theory between bosons and fermions. 
For every fermion degree of freedom, supersymmetry imposes that there must be
a corresponding bosonic one, with identical mass and couplings. Thus, for
instance in Fig.~\ref{hierProb}, supersymmetry imposes $m_F=m_\varphi$ and
$\lambda^2=\xi$~\cite{Quevedo:2010ui}. 

In fact, fermions in the Standard Model (the quarks and leptons), each have
two degrees of freedom (left and right handed, meaning that their spins are in
the same 
direction or in opposite direction to the motion of the particle). When we
supersymmetrise the 
model, we end up 
with two scalar bosons for each Standard Model fundamental fermion. The
supersymmetric scalar boson copies are prepended with an `s' to denote their
different spin. Thus, we talk of right and left handed squarks and sleptons to
be the spin 0 copies of the fermions. In terms of representation of the
Standard Model gauge group $SU(3) \times SU(2)_L \times U(1)_Y$, we have, in
the minimal supersymmetric extension:
\begin{itemize}
\item (s)quarks: lepton number $L=0$, whereas baryon number $B=1/3$ for a
  (s)quark, $B=-1/3$ for an anti-quark.
    \[{Q_{i} = \left(3 , \ 2 , \tfrac{1}{6}
      \right)} ,\qquad {{u}_{i}^{c} =
      \left(\bar{3} , \ 1 , \ -\tfrac{2}{3} \right) ,\qquad {d}_{i}^{c} =
      \left(\bar{3} , \ 1 , \ \tfrac{1}{3} \right)}
\]
\item (s)leptons $L=1$ for a lepton, $L=-1$ for an anti-lepton. $B=0$.
    \[{L_{i} = \left(1 , \ 2 , \ -\tfrac{1}{2}
      \right)} ,\qquad {{e}_{i}^{c} = \left(1
      , \ 1 , \ +1 \right) }
\]
\item higgs bosons/higgsinos: $B=L=0$.
    \[H_{2} = \left(1 , \ 2 , \ \tfrac{1}{2} \right) ,\qquad H_{1} = \left(1 , \ 2 , \ -\tfrac{1}{2} \right)
\]
\end{itemize}
the second of which is a new Higgs doublet not present in the Standard
Model. Thus, the minimal supersymmetric standard model (MSSM) is a {\em two Higgs doublet model}.
The extra Higgs doublet is needed for mathematica consistency (i.e.\ in order
to avoid a $U(1)_Y$ gauge anomaly), and to give
masses to down-type quarks and leptons. $B$ and $L$ denote baryon number and
lepton number, respectively. $B$ and $L$ are usually assumed to be conserved
in perturbative interactions (except in so-called $R$-parity violating
models~\cite{Barbier:2004ez}). For the rest of this article, we shall take the
assumption that $R-$parity is conserved, and therefore that $B$ and $L$ are
separately conserved. 

The spin 1 force-carrying gauge bosons of the Standard Model (prior to the
Higgs mechanism, these are gluons, $SU(2)$
bosons and hypercharge gauge bosons, respectively) attain a spin 1/2
supersymmetric partner, collectively known as {\em gauginos}. Their
Standard Model representations are:
\begin{itemize}
\item gluons/gluinos
\[ G = (8, 1, 0)\]
\item
$W$ bosons/winos
\[ W = (1, 3, 0)\]
\item
$B$ bosons/gauginos
\[ B =(1, 1, 0), \].
\end{itemize}

The supersymmetric prediction that, for instance, the slepton masses are
identical to the lepton masses leaves us with a phenomenological problem. 
To date, no supersymmetric particles have been directly observed. But if they
were of identical mass to their Standard Model counterparts, they would have
been observed in past experiments. The resolution to this problem is
the introduction of {\em supersymmetry breaking} in a way that does not
reintroduce the hierarchy problem. 
Supersymmetry breaking which does not reintroduce the hierarchy problem is
called soft. If we introduce a mass splitting between $F$ and $\varphi$, we
induce a correction to the Higgs mass squared of order
\begin{equation}
\frac{1}{16 \pi^2} (m_F^2 - m_\varphi^2).
\end{equation}
Thus, no matter the size of $m_F$, as long as the {\em splitting}~with its
partner is small, there is no large quantum correction induced to the Higgs
mass. The mass of the Higgs then gives us a rough order of magnitude estimate
for what the splitting should be: it should not be too much greater than the
Higgs mass times 4$\pi$ (i.e.\ approximately 1000 GeV). Thus, the masses of
the supersymmetric partners of the Standard Model particles (which only have a
comparatively negligible mass)
should not be too much greater than about 1000 GeV. 
Many different models have been suggested
that successfully exhibit such soft supersymmetry breaking, conveniently
making 
supersymmetric partners heavier while not significantly affecting the masses
of the Standard Model particles. Although there are many different models,
with different predictions for the patterns of masses of supersymmetric
particles and slightly different advantages or disadvantages, there is no
outstanding candidate. In any case, we are faced with the hope that
supersymmetric particles will be produced at the LHC, and measurements of
their masses and couplings will subsequently be made, allowing for an empirical
determination of the pattern of supersymmetry breaking. 

In order to deduce what signatures are expected in LHC collisions which
produce supersymmetric particles, we must examine the model {\em after}
supersymmetry breaking and electroweak symmetry breaking via the 
Higgs mechanism~\cite{beh}. Various particles
mix after the effects of breaking these two symmetries are taken into account.
For example, the electrically neutral spin 1/2 particles with $L=0$ (i.e.\ the
Higgsinos, the Zino and the photino) mix: their mass eigenstates are called
neutralinos, and denoted $\chi_{1,2,3,4}^0$. Neutralinos can help solve
another problem associated with cosmological and astrophysical observations of
our universe: namely, the dark matter problem. 

\begin{figure}[!h]
\centering\includegraphics[width=3in]{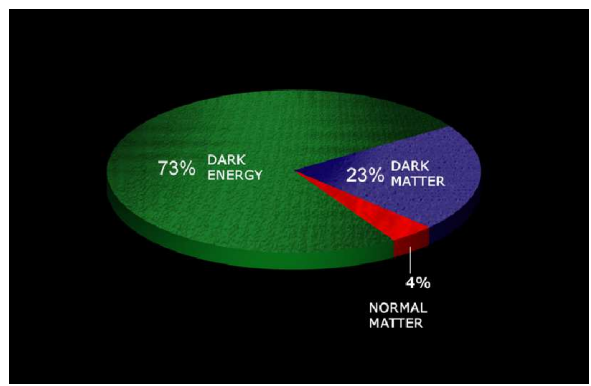}
\caption{Energy budget of the universe: there is approximately six times the
  amount of mass in dark matter compared to normal matter. The rest of the
  energy is in the mysterious `dark energy' of the universe, about which very
  little is known.}
\label{fig:pie}
\end{figure}

Dark matter is a hypothesised new form of particle which is transparent to
light, and which only interacts very weakly with matter. However, it is heavy
and can affect the gravitational field in the universe. It was initially
postulated to be present in a halo around spiral galaxies, because the
speeds of the stars rotating around their centres did not match up with the
predictions coming from standard gravitational theory (the stars more toward
the edges of the galaxies were going far too fast). By postulating some
heavy invisible matter hanging around though, the predictions changed and the
speed of the outer stars in such galaxies could be understood. Other
corroborating inferences from observations soon were made: clusters of
galaxies were seen to be moving with
peculiar velocities that could be explained if dark matter were
present. Also, measurements of the bending of light (weak gravitational
lensing) from distant galaxies 
indicate that the light has passed through dark matter.
Recently, observations including those of the afterglow of the big bang
(effectively the angular correlations in the temperature spectrum of the
cosmic microwave background) allow a fit in order to determine the amount of
dark matter in the universe, compared to the amount of visible matter. While
ordinary 
(baryonic) matter only makes up around 4$\%$ of the energy budget of our
universe, dark matter makes up some $23\%$ or so (see Fig.~\ref{fig:pie}.)

The lightest neutralino,
$\chi_1^0$ can have the right properties to make up the dark matter of the
universe, since it is massive and does not interact with light. For this to be
the case, it must be stable, which in 
practice means that it must be the lightest supersymmetric particle
(supersymmetric particles can only decay to another lighter supersymmetric
particle and ordinary Standard Model particles). 
This then leads to a rewriting of the early universe's history: in the first
instants after the big bang, the universe is very hot and energetic. There are
all sorts of particles, including various supersymmetric ones. The
supersymmetric particles all then very quickly decay away into ordinary
(Standard Model)
particles and dark matter, which hangs around in the universe to this day. 
The idea then, is that although the supersymmetric particles (aside from the
dark matter) have all long since decayed, we can convert the energy $E$ in the
proton beams of the LHC into mass $m$ of supersymmetric particles through the
famous relation $E=mc^2$, so that we
can measure some of their decay products and confirm their existence.

\section{Universality and Large Hadron Collider Searches}
\begin{figure}
\centering\includegraphics[width=3in]{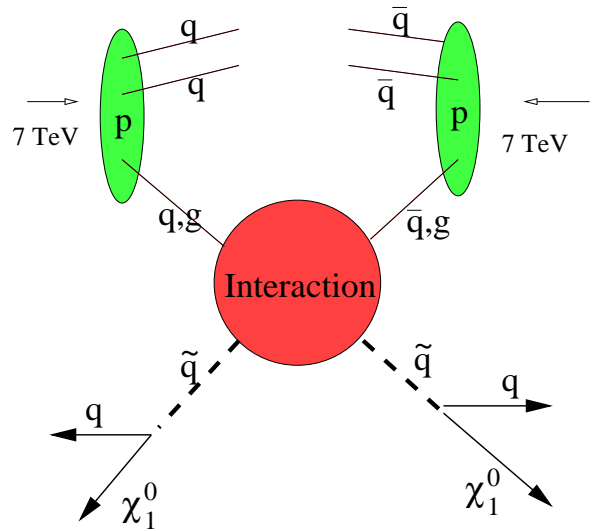}
\caption{Sketch of a collision producing supersymmetric particles at the
  LHC. The initial protons of the beams (denoted `p`) are made of quarks (`q')
and gluons (`g'). It is these which may collide to produce supersymmetric
particles (these are pair produced, and shown as $\tilde q$). The
supersymmetric particles subsequently decay to ordinary particles and dark
matter particles $\chi_1^0$.}  
\label{fig:coll}
\end{figure}
The preceding decades have seen many different colliders with successively
higher and higher energies. Since the 1960's, every decade the most energetic
man-made particle collisions on earth have had their energies increased by
roughly a factor 10. During this period, as the energy was increasing, various
particles were discovered. $E$ was not high enough in
previous experiments in order to produce the heavier states (which have a
larger $m$). Today, the LHC has the highest energy of any artificial particle
collider that has ever existed on earth. Thus, if $m$ is significantly higher
than the 
$E$ of any of the previous collisions, but still somewhat less than the LHC
design energy of 
14000 GeV, we would expect to be able to firs produce them at the LHC,
provided that nature
follows the supersymmetric model. We argued above that the supersymmetric
particles should have masses less than 1000 GeV or so.
This in turn implies that
the Large Hadron Collider, with its centre of mass energies, at 7000 GeV-14000
GeV, should be energetic enough to produce the supersymmetric partners. We
must bear in mind though that, in fact, it is the point-like constituents of
the protons (the quarks and gluons) that may collide to produce
supersymmetric partners, and they come with some random fraction of the
proton's energy in each collision (see Fig.~\ref{fig:coll}). A
combination of increasing the 
energy and recording more collisions allows for the greatest chance of directly
producing the supersymmetric partners. The hope is that, after production,
they can 
be detected and measured in the ATLAS and CMS general purpose LHC detectors. 
These large machines act like three dimensional digital cameras, measuring the
properties of the fiery fragments (momenta, charges etc) resulting from the
proton collisions. Some detective work is required to work backwards from the
tracks of the fiery fragments and tell what is happening directly after the 
moment of collision. We must bear in mind that the collisions are quantum
processes 
and their outcomes are inherently random. One cannot predict, for a given
collision, what will result. However, if one uses the correct quantum theory
to describe the collisions, one can predict the relative probabilities of
various possible final states of a collision. Thus, after performing many
collisions, one can check the predicted frequencies of various final states
against the measured frequencies. These should match within uncertainties, if we
have the correct 
theory, provided there have been no mistakes in the analysis. 

The classic signature for supersymmetry in the LHC is that of missing
transverse momentum. School-level physics tells us that momentum is conserved
in any LHC collision. Initially, we have protons of equal and opposite
momentum, and so the total momentum of the initial state is 0. The law of
conservation of momentum implies that the vector sum of the final state
momenta should therefore also be zero. However, if we add up the components of
the momenta transverse to the beam (there are additional difficulties with
measuring the total momentum parallel to the beam which renders it
impractical) and there is a significant amount `missing', we can infer one of
two things: either some particles fell in cracks in the detector and were
unmeasured, or a particle went right through the detector without leaving a
trace. One 
can account (by careful modelling and callibration) for the former
effect. Each supersymmetric 
particle would decay to some ordinary particles and a single dark matter
particle. The dark matter particle interacts so weakly with matter that it
would just go straight through the detector without leaving a trace. It thus
acts like a thief, stealing momentum away from the collision, undetected. If
we measure too many collisions which have a large amount of `missing energy',
we can infer the production of such particles. By measuring some of the
details of the visible particles that are produced, we can hope to check
aspects of the supersymmetric model. 

\begin{figure}
\centering\includegraphics[width=5in]{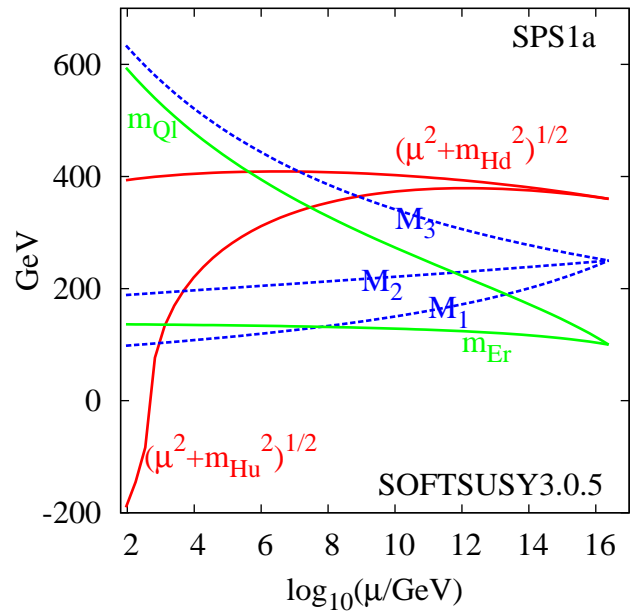}
\caption{Example of quantum corrections splitting the various supersymmetric
  particle masses. On the right-hand side of the plot, the spin 0 particle
  masses $m_{Er}$ and $m_{Ql}$ are equal. As the model is evolved down to
  lower energies $\mu$ relevant 
  for experiments, the masses split apart due to differing quantum
  corrections. The same can be said of the gaugino masses $M_1, M_2,
  M_3$. Obtained by the publicly available {\tt SOFTSUSY}~\cite{softsusy}
  program. The parameter point (SPS1a) is defined in
  Ref.~\cite{Allanach:2002nj}.}  
\label{fig:renorm}
\end{figure}
Unfortunately, the different ways in which supersymmetry can be broken leads
to many different possible patterns of the details of the final fiery
fragments of the 
collisions. For a particular pattern of supersymmetry breaking, we can make
predictions for the various relative frequencies of the possible final
states. In order to provide a realistic example, we often resort to {\em
  universal} models of supersymmetry breaking. These assume that all of the
supersymmetry breaking spin 0 particle masses are equal ($m_0$), all of the
supersymmetry breaking gaugino masses are equal ($M_{1/2}$), and that each of
the supersymmetry breaking trilinear interactions between spin 0 particles are
equal ($A_0$). There is another input parameter in the theory, $\tan \beta$,
which measures how different the two Higgs doublet in the model are. Once
these four parameters are set (and a sign in the Higgs potential, the sign of
the so-called $\mu$ parameter), the model can be matched to current data on
Standard Model particle masses, and the masses of all supersymmetric particles
and Higgs' can be predicted. In fact, the initial equal masses for the
supersymmetric particles split apart because of differing quantum corrections:
see Fig.~\ref{fig:renorm}.
We shall assume the universal pattern of supersymmetry breaking
throughout the present article. 

\begin{figure}
\centering\includegraphics[width=5in]{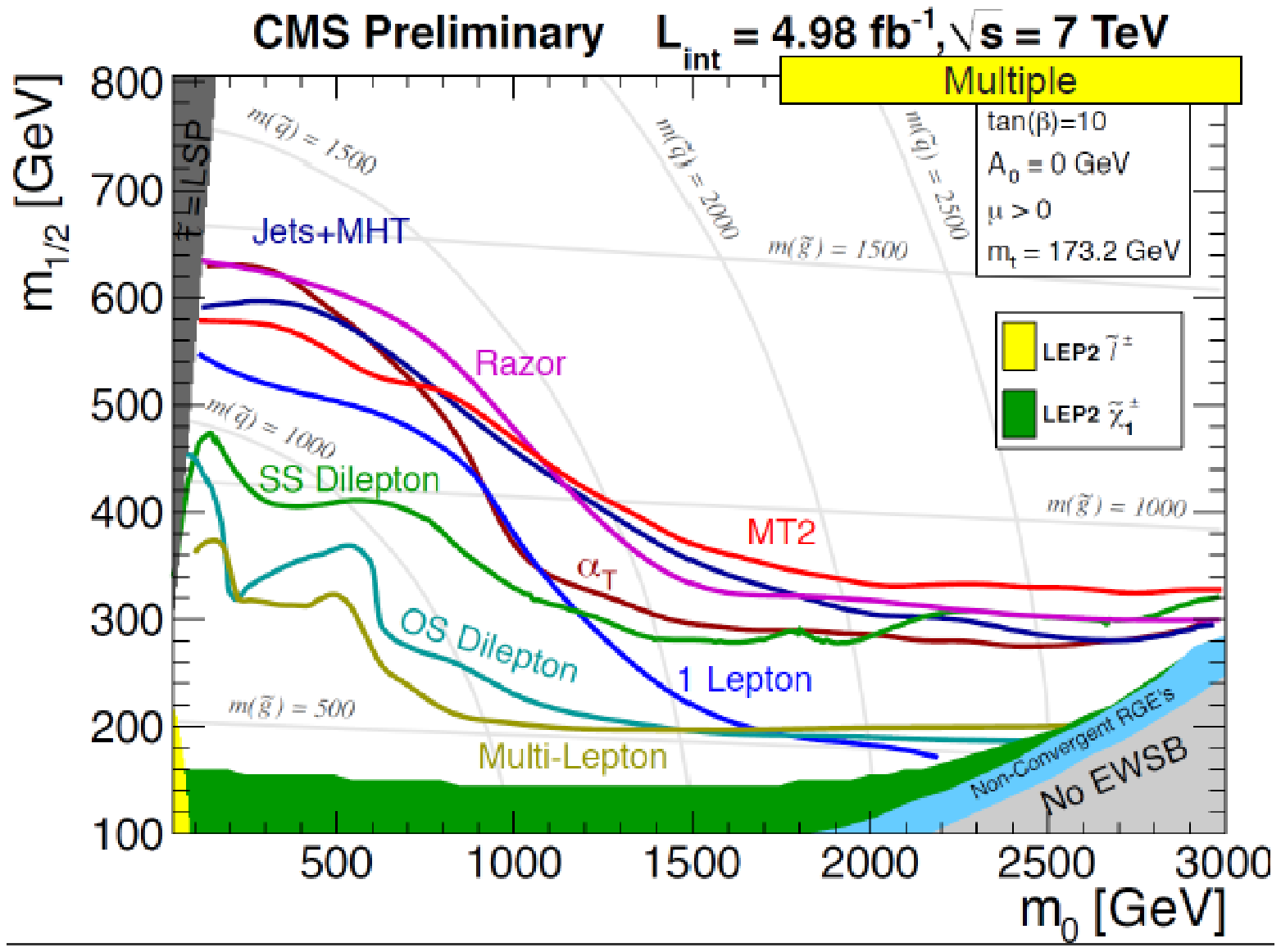}
\caption{Exclusion limits from different experimental searches carried out at
  a centre of mass energy of 7 TeV at the LHC. Each curve shows exclusion
  limits derived from 
  a
  different search. For each point in this parameter plane, the properties
  (such as masses) of the supersymmetric particles are set, although they vary
  from point to point. Each of the searches yielded a null result and so the
  area under each curve is ruled out by the corresponding search. The number
  of curves illustrates the many 
different ways in which supersymmetry is searched for at the LHC.}  
\label{fig:limits}
\end{figure}

So far, no direct evidence for supersymmetric particle production at the LHC
has been found (in other words, not enough collisions have been seen which
predict high amounts of missing transverse momentum).
In order to display this fact, experiments interpret their data
in terms of exclusion bounds on supersymmetric models. For example, in
Fig.~\ref{fig:limits}, we see exclusion bounds on the universal model
described above. As can be seen from the figure, are many ways of sieving the
data and looking for supersymmetric particles, with much associated activity
by 
the experimental collaborations in doing so. 

\section{Multiple Solutions}
\begin{figure}
\centering\includegraphics[width=5in]{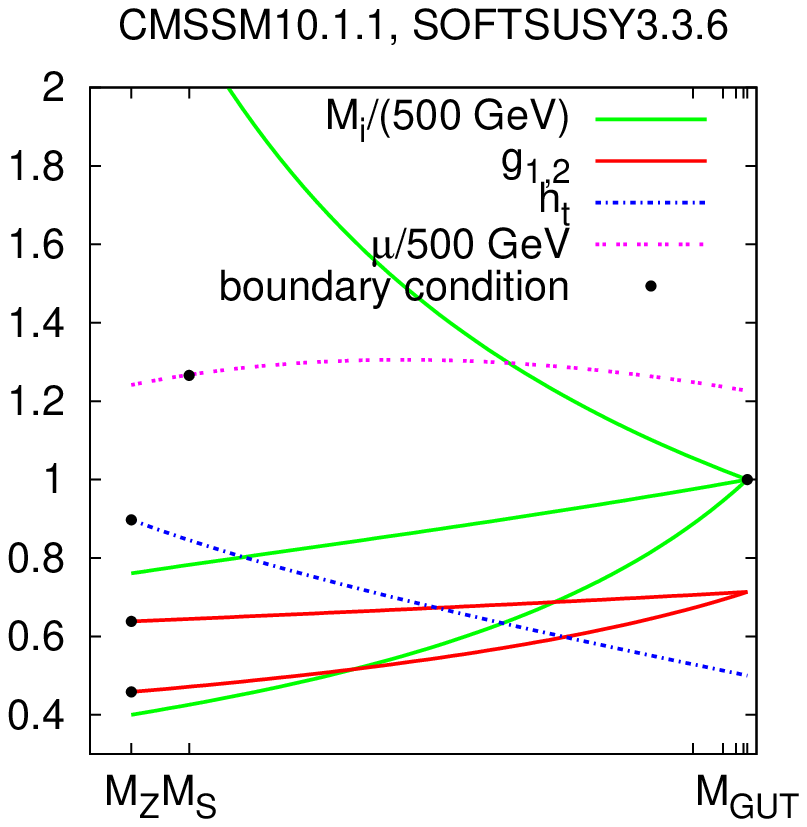}
\caption{Boundary conditions in universal supersymmetry breaking models. We
  have shown the evolution of some of the mass parameters as the curves. The
  points show where boundary conditions are imposed: At $M_Z$, we impose those
  coming from   experimental measurements of Standard Model particle masses
  and couplings, at $M_S$ boundary conditions coming from minimising the Higgs
potential are imposed, and at $M_{GUT}$, theoretical boundary conditions on
the supersymmetry breaking terms are imposed. From
Ref.~\protect\cite{Allanach:2013cda}.}   
\label{fig:multiBCs}
\end{figure}
In order to interpret the various supersymmetric particle searches, one must
solve the differential equations that dictate the behaviour of the
supersymmetric particle masses. One such solution is shown in
Fig.~\ref{fig:renorm}. Such 
differential equations are shown by the Cauchy-Lipschitz theorem to have
a unique solution, provided certain conditions hold (Lipschitz continuity, and
that the boundary condition is fixed at one point; in this case at $M_{GUT}
\sim 10^{16}$ GeV). In practice though, one of these conditions is violated:
since actually, the boundary conditions are set at radically different
points: $\mu=M_{GUT}$, $\mu=M_Z=91$ GeV and $\mu=M_{S}\sim$1000 GeV. The
problem is then a {\em boundary value problem}\ rather than an {\em initial
  value problem}. A cartoon of the situation is shown in
Fig.~\ref{fig:multiBCs}.  

In a recent publication~\cite{Allanach:2013cda}, it was shown how the multiple
boundary 
conditions allow several solutions to the system of boundary conditions and
differential equations. These multiple solutions have some parameters being
different, and therefore have different associated spectra. In principle, the
different 
spectra can lead to different predictions for the outcome of collisions at the
LHC. This leads to a potential loophole in interpretations of data, such as
those in Fig.~\ref{fig:limits}: {\em if one does not know of the existence of
  the 
  additional solutions, one could be ruling out a point in parameter space
  from interpreting the data where one should not}.
It was decided to investigate the properties of the additional solutions, and
determine if they could cause such loop-holes. New techniques had to be used
to find the multiple solutions, because they are unstable to the usual 
algorithm for solving the system
(fixed point iteration). In Ref.~\cite{Allanach:2013yua}, we showed that the
shooting method 
(with plenty of `shots') can solve the problem. 

\begin{figure}\begin{center}
\includegraphics[width=0.8\textwidth]{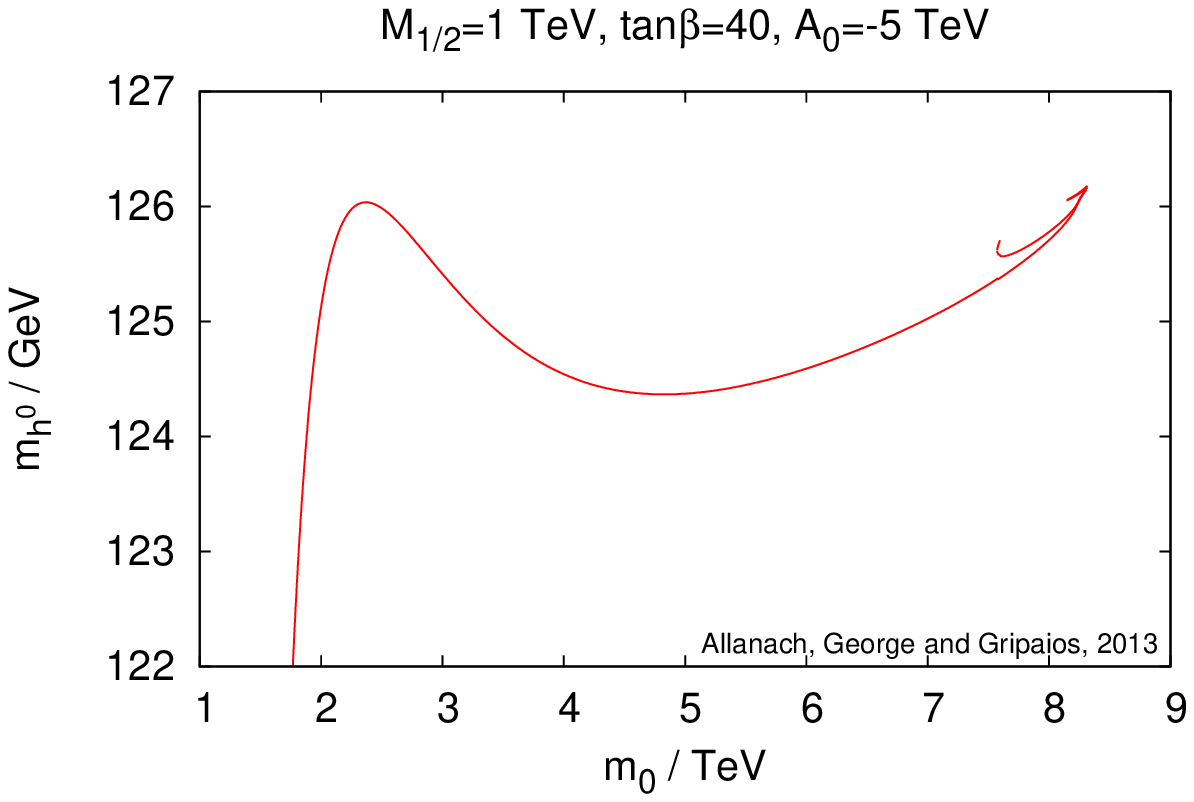}
\caption{Multiple branches of solutions in the universal model at $M_{1/2}=1$~TeV,
  $A_0=-5$~TeV and $\tan\beta=40$. The value of $m_0$ ranges as per the horizontal
  axis.  We plot here the predicted values of the Higgs mass, $m_{h^0}$.  There are
  3 solutions in the range $7572\;\text{GeV} \leq m_0 \leq 7595\;\text{GeV}$
  which are consistent with recent LHC measurements of a Higgs boson
  mass. Figure taken from Ref.~\cite{Allanach:2013cda}.
\label{fig:multiPheno-mh0vsm0}}
\end{center}
\end{figure}
Here, we shall illustrate some of the properties of the multiple solutions. 
For instance, in Fig.~\ref{fig:multiPheno-mh0vsm0}, we show the predictions
for the Standard Model-like Higgs mass for a particular parameter choice,
allowing the universal scalar mass $m_0$ to vary. 
$m_0>2000$ GeV is 
consistent with recent LHC measurements of
a Higgs boson~\cite{ATLAS:2012oga,Chatrchyan:2012ufa}.  For
$m_0=7572-7595$ GeV\footnote{As is well known, getting $m_h \sim 125$
  GeV in the MSSM in general requires unnaturally large stop quark masses, and
  hence large $m_0$ in the universal model.}
there are 3 solutions, 2 with $\mu(M_S)<0$, which have $m_{h^0}$ in
the range $125.4$ GeV to $125.7$ GeV. We show another parameter point in
Table~\ref{tab:spec}. 
\begin{table}
\begin{center}
\begin{tabular}{|c|ccc|}\hline
quantity & solution A & solution B & solution C \\ 
\hline 
\hline 
$M_{\chi_1^0}$/GeV    & 282 & 282 & 281 \\
$M_{\chi_2^0}$/GeV    & 502 & 497& 471 \\
$M_{\chi_3^0}$/GeV    & 558 & 548 & 510 \\
$M_{\chi_4^0}$/GeV    & 610 & 605 & 593 \\
$M_{\chi_1^\pm}$/GeV & 503& 497& 470 \\
$M_{\chi_2^\pm}$/GeV & 609 & 604 & 592 \\
$m_{\tilde g}$/GeV    & 1612 & 1612 & 1612  \\ \hline
$\mu(M_S)$/GeV & -545 & -535& 497 \\ 
$m_3^2 (M_S)/10^5$ GeV$^2$ & 0.800 & 0.809 & 1.07 \\
$m_{H_2}^2 (M_S)/10^5$ GeV$^2$ & -1.94 & -1.83& -1.42 \\
$h_t(M_S)$ & 0.840& 0.839 & 0.836 \\
$A_t(M_S)$/GeV & -1056&-1057 & -1064 \\
$M_X/10^{16}$ GeV & 1.94 & 1.93& 1.89 \\
$g_1(M_Z)$ & 0.460 & 0.470& 0.456 \\
$g_2(M_Z)$ & 0.634 & 0.640& 0.633 \\ 
\hline\end{tabular}
\caption{\label{tab:spec} Differences in universal parameters and spectra for the
  multiple 
  solutions of the parameter point $m_0=2.8$ TeV, $M_{1/2}=660$ GeV, $\tan
  \beta=40$ and $A_0=0$. 
  The solutions are found by scanning $\mu(M_S)$ and then the rest of the
  quantities are determined by the iterative algorithm.
  We display here some masses and parameters of interest for the 3 solutions
  that predict the correct value of $M_Z$.
  The standard technique (fixed point iteration) only finds solutions B and
  C. The first 
  quantities listed are selected supersymmetric particle masses, whereas those
  under the horizontal lines are some underlying parameters of the model that
  are fixed by the boundary conditions. 
  Figure taken from
  Ref.~\cite{Allanach:2013cda}.} 
\end{center}
\end{table}
The spectra show some notable
differences, illustrating the fact that the solutions are physically
different, leading to the possibility of their discrimination by collider
measurements. Masses whose tree-level values depend upon the value of
$\mu$, such as the heavier neutralino and chargino masses, 
show the largest differences. Other
sparticle and Higgs masses do have small per-mille level fractional
differences for this parameter point.

\begin{figure}
\begin{center}
\unitlength=\textwidth
\begin{picture}(1,0.5)
\put(0,0){\includegraphics[width=0.5\textwidth]{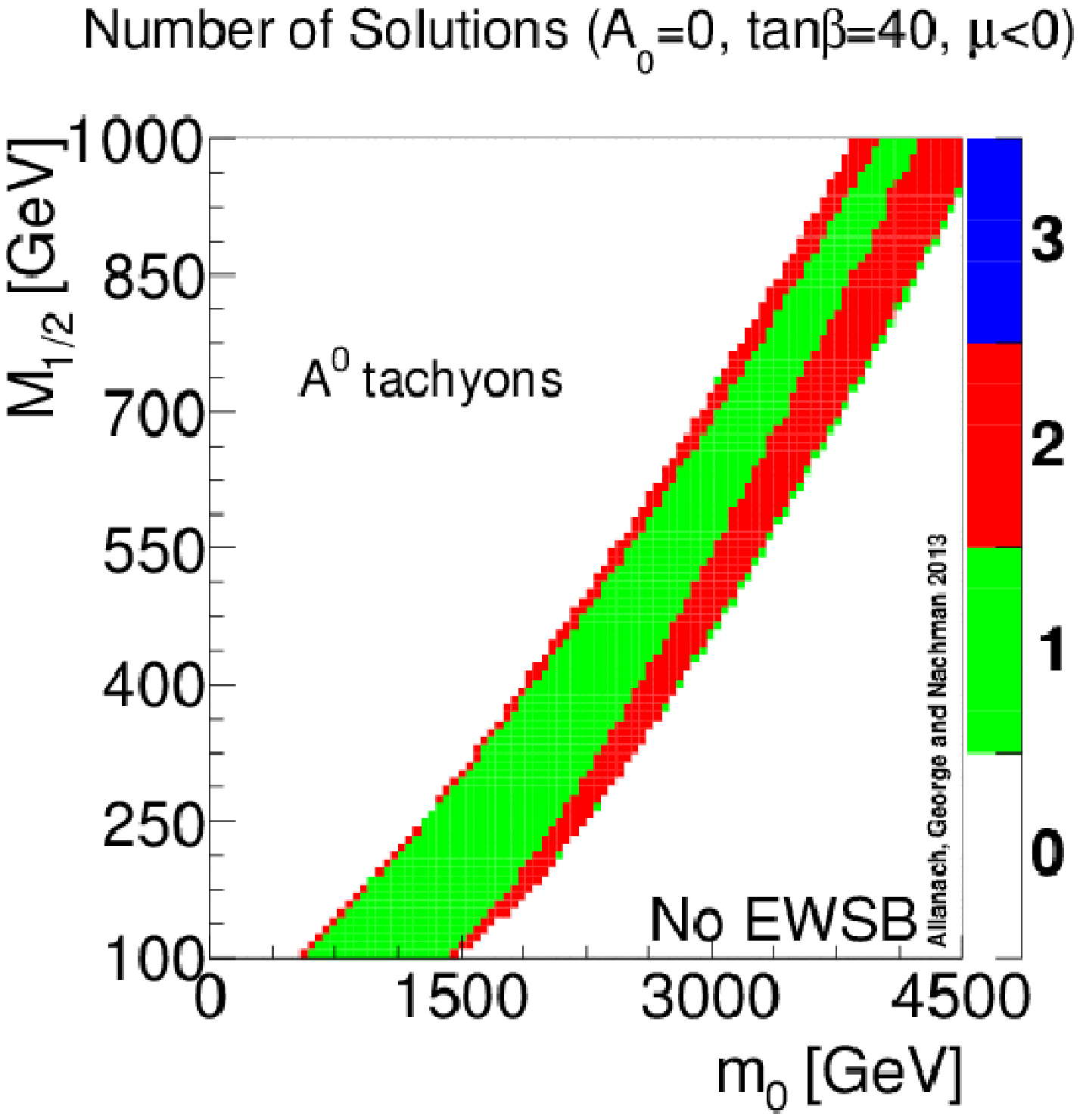}}
\put(0.5,0){\includegraphics[width=0.5\textwidth]{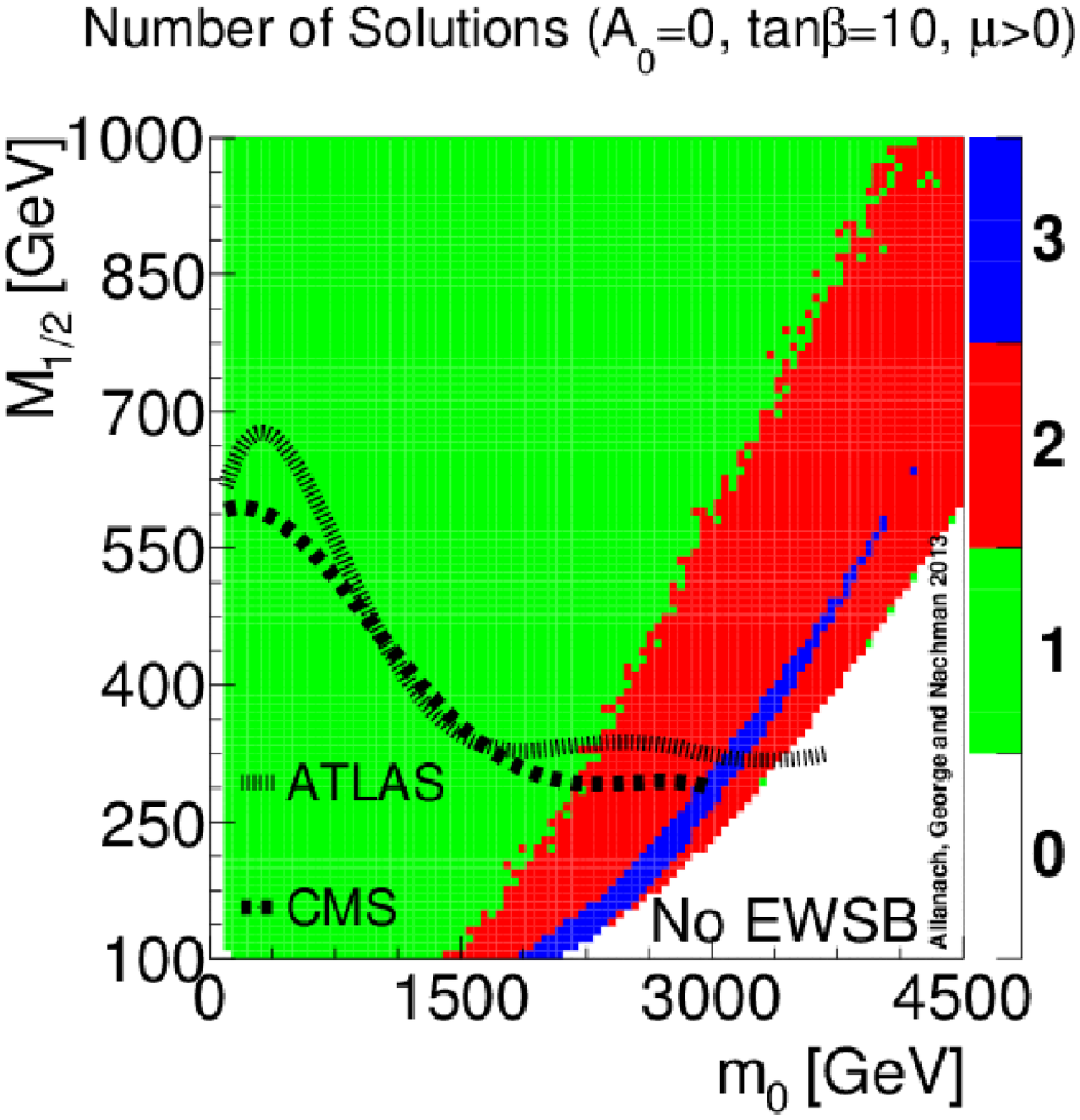}}
\put(0,0.5){(a)}
\put(0.5,0.5){(b)}
\end{picture}
\end{center}
\caption{\label{fig:numberofsolutions} Number of solutions in the universal
  model  as
  shown as the background colour and labelled in the key on 
  the right-hand side of each plot. White
  regions have no solutions for the reasons labelled: `No EWSB' denotes a
  region where there is no acceptable electroweak minimum of the Higgs
  potential. The lines in (a) display 95$\%$ exclusion contours
  from ATLAS \cite{Aad:2012fqa} and CMS~\cite{Chatrchyan:2012lia} jets plus
  missing transverse momentum searches. The region below
  each contour is excluded. Figure taken from Ref.~\cite{Allanach:2013yua}.}
\end{figure}
We show the number of solutions found along a particular parameter plane in
Fig.~\ref{fig:numberofsolutions}a. We have chosen $A_0=0$, $\tan \beta=10$ and
$\mu>0$ in
particular because the LHC experiments ATLAS and CMS have interpreted their
most sensitive searches in terms of exclusion regions with those values of the
parameters. We
see that their excluded regions include points where up to 3 solutions are
predicted. In fact, it turns out that in this plane, all of the multiple
solutions have already been ruled out by previous experimental searches for
charginos. Thus they do not present a problem. 

At higher values of $\tan \beta$ and $\mu<0$, there is a strip where this is
not a problem, and is shown in Fig.~\ref{fig:numberofsolutions}b: the
uppermost strip with two solutions (the lower strip is ruled out by having
light charginos that would have been seen at the LEP2 collider). In fact, the
multiple solutions in Table~\ref{tab:spec} are taken from this
`phenomenologically plausible strip'. Squarks from the first two generations
and gluino masses display a negligible difference between the different
solutions in this strip. The neutralino mass can be different by about
1-2$\%$, depending upon the position in the plane. This means that, for the
most stringent searches involving multiple 
jets of hadrons and missing transverse energy, if one of the solutions is
ruled out by the analysis, the other solution will also be ruled out. On the
other hand, chargino masses can be affected by 10$\%$ or so, and so
interpretations of analyses which depend upon charginos in decay chains may be
sensitive to the multiple solutions, and should be checked on a case-by-case
basis. 

Here, we exemplify the most important difference between the solutions: that
of a different predicted thermal relic density of dark matter in the universe.
Recently, data from the Planck satellite have been used to derive the
constraint~\cite{Ade:2013zuv} on the thermal dark matter relic density
\begin{equation}
\Omega_{CDM} h^2=0.1198 \pm 0.0026.
\end{equation}
We place a dominant
theoretical uncertainty on our prediction of 0.01 coming from
loops (the thermal relic density is only calculated by the publicly available
{\tt 
  micrOMEGAs}~\cite{Belanger:2006is,Belanger:2008sj} program to
tree-level order), and therefore require the predicted 
thermal relic density of neutralinos to be 
$\Omega_{CDM} h^2 \in [0.0998,\ 0.1398]$.
After a brief scan, we found a 
parameter point in the phenomenologically plausible strip ($m_0=760$ GeV, $M_{1/2}=141.72$ GeV, $A_0=0$,
$\tan \beta=40$, $\mu<0$)  
where the standard solution predicts $\Omega_{CDM} h^2=0.34$, i.e.\ well outside of
this range, but 
where the {\em additional} solution prediction of $\Omega_{CDM} h^2=0.118$ is near
the central value. It turns out that this point has the $\chi_1^0$ mass being
approximately half of the Higgs mass. The $\chi_1^0$ mass, which changes
slightly between the solutions, is more exactly half the lightest CP even
Higgs mass for the 
additional solution, which leads to very efficient annihilation of neutralinos
through an $s-$channel Higgs boson into quark or lepton pairs, significantly
reducing the relic density from 0.34 to 0.118.

\section{Summary}
Supersymmetry is a well-motivated theory that explains how the Higgs
mass is insensitive to potentially huge quantum corrections. It predicts a
gamut of new particles with specific properties, that are being actively
searched for at the Large Hadron Collider. In the absence of a signal of the
production of supersymmetric particles (given simply by an excess of
collisions in which there is a large apparent missing transverse momentum), 
the data are interpreted in terms of exclusion limits on models of
supersymmetry breaking. Such exclusion limits have a potential loop-hole, due
to the existence of multiple solutions, which have only just been found
recently in the literature. Each limit, which was interpreted only as a single
solution, should be checked to see if it changes when the multiple solutions
are taken into account. We have checked in universal supersymmetry breaking
models that the most stringent searches -
involving jets of hadrons and missing transverse energy - are insensitive to
the multiple solutions, for several reasons. Either the multiple solutions are
ruled out by previous experiments because they predict very light charginos,
or the masses of lightest neutralinos and squarks and gluinos are similar
enough between the solutions such that if one solution is ruled out by an
analysis, to a good approximation the other solution will be as well. However,
more particular searches such as those involving searches for charginos are
likely to display significant differences between solutions, and the limits
must be interpreted carefully for each one in turn. Also, we have demonstrated
with a parameter point example that the predicted density of dark matter left
today in the universe can be very sensitive. The parameter point had far too
much dark matter in the standard solution compared to the amount derived from
observations, whereas the additional solution had just the right amount. 
Analyses employing the relic density of dark matter can therefore be
particularly 
vulnerable to changes coming from the existence of additional solutions.
We expect multiple solutions as a possibility whenever a high-scale
supersymmetry breaking mechanism is active, such as in superstring-inspired
models. 

\section*{Acknowledgment}
This work has been partially supported by STFC. We thank the Cambridge SUSY
Working Group for helpful discussions.


\end{document}